\documentclass[
amsfonts, 
amssymb, 
amsmath, %
floatfix, %
twocolumn, %
reprint, %
superscriptaddress, %
prb, %
aps, %
nofootinbib,%
nobibnotes,%
]{revtex4-1}
\usepackage[T1]{fontenc}
\usepackage[utf8]{inputenc}
\usepackage[]{graphicx} 
\usepackage{latexsym}
\usepackage{amsthm}

\usepackage{color}
\usepackage{mathtools}
\usepackage[usenames]{xcolor}
\usepackage{bm}

\usepackage{multirow}

\usepackage{floatrow}
\usepackage{hyperref}
\hypersetup{
	colorlinks,
	linkcolor={blue!90!black},
	citecolor={blue!90!black},
	urlcolor ={blue!90!black}
}
\makeatletter
\renewcommand*{\eqref}[1]{%
	\hyperref[{#1}]{\textup{\tagform@{\ref*{#1}}}}%
}
\makeatother
\newcommand{\onlineciteb}[1]{[\onlinecite{#1}]}

\usepackage{siunitx}
\newcommand{\rr}{\bm{r}}
\newcommand{\RR}{\bm{R}}
\newcommand{\kk}{\bm{k}}
\newcommand{\qq}{\bm{q}}

\newcommand{\sumql}{\sum_{\bm{q},\lambda}}
\newcommand{\sumq}{\sum_{\bm{q}}}

\newcommand{\wql}{\omega_{q,\lambda}}

\newcommand{\bql}{b_{\bm{q},\lambda}}
\newcommand{\bqld}{b_{\bm{q},\lambda}^{\dag}}
\newcommand{\bmqld}{b_{-\bm{q},\lambda}^{\dag}}
\newcommand{\eql}{\hat{\bm{e}}_{\bm{q},\lambda}}
\newcommand{\emqlc}{\hat{\bm{e}}_{-\bm{q},\lambda}^{*}}
\newcommand{\kp}{\mbox{$\bm{k}\!\cdot\!\bm{p}$ }}

\newcommand{\cp}{\mathrm{c.p.}}
\newcommand{\hc}{\mathrm{H.c.}}
\newcommand{\es}{\epsilon^{(\mathrm{s})}}
\newcommand{\eph}{\epsilon^{(\mathrm{ph})}}

\newcommand{\tcal}{\mathcal{T}}

\newcommand{\Eg}{E_{\mathrm{g}}}
\newcommand{\Eso}{E_{\mathrm{g}} +\Delta_{\mathrm{SO}}}

\newcommand{\lr}[1]{\!\left(#1\right)} 
\newcommand{\Lr}[1]{\!\left[#1\right]}

\newcommand{\hf}{\frac12}

\DeclareMathOperator{\tr}{Tr}
\newcommand{\tabtopspacer}{\rule{0pt}{2.4ex}}

\newcommand{\minus}{\mkern1mu{-}\mkern1mu}
\newcommand{\plus}{\mkern1mu{+}\mkern1mu}

\begin{document}

\title{Dominant role of shear-strain-induced admixture in spin-flip processes in
  self-assembled quantum dots}  
\author{Adam Mielnik-Pyszczorski}
	\affiliation{Department of Theoretical Physics, Faculty of Fundamental
	Problems of Technology, Wroc{\l}aw University of Science and
	Technology, 50-370 Wroc{\l}aw, Poland}

\author {Krzysztof Gawarecki}
	\affiliation{Department of Theoretical Physics, Faculty of Fundamental
		Problems of Technology, Wroc{\l}aw University of Science and
		Technology, 50-370 Wroc{\l}aw, Poland}

\author{Micha{\l} Gawełczyk}
	\affiliation{Department of Theoretical Physics, Faculty of Fundamental
	Problems of Technology, Wroc{\l}aw University of Science and
	Technology, 50-370 Wroc{\l}aw, Poland}
	\affiliation{Department of Experimental Physics, 
          Faculty of Fundamental Problems of Technology, 
          Wroc\l{}aw University of Science and Technology, 50-370 Wroc\l{}aw, Poland}

\author{Pawe{\l} Machnikowski}
	\email{Pawel.Machnikowski@pwr.edu.pl}
	\affiliation{Department of Theoretical Physics, Faculty of Fundamental
 	Problems of Technology, Wroc{\l}aw University of Science and
 	Technology, 50-370 Wroc{\l}aw, Poland}

\begin{abstract}
We study theoretically the spin-flip relaxation processes for a single electron in a self-assembled
InAs/GaAs quantum dot and show that the dominant channel is the spin admixture induced by
symmetry-breaking shear strain. This mechanism, determined within the 8-band envelope-function
\kp theory, can be mapped onto two effective spin-phonon terms in a conduction-band (effective
mass) Hamiltonian that have a similar structure and interfere constructively. Unlike the
Dresselhaus coupling, which dominates spin relaxation in larger, unstrained dots, the shear
strain contribution cannot be modeled by a generic, standard term in the Hamiltonian but
rather relies on the actual strain distribution in the quantum dot. 
\end{abstract}


\maketitle

\section{Introduction} 
\label{sec:intro}

Dynamics and decoherence of spins in quantum dots (QDs) has been in
the focus of  both experimental and theoretical studies for several
years. This research activity is motivated by the scientific
interest in this non-trivial and still not fully understood problem,
as well as the promise it holds for possible applications in
spintronics and quantum information processing
\cite{loss98,recher00}. The latter is fed by 
the experimental results showing very long life times of confined
spins that raise hopes for their applications as spin memories
\cite{kroutvar04} and by the development of manufacturing and
control technologies that allow one to coherently drive quantum
spin states in a desired way \cite{koppens06,Nowack_Science07}. 

Among various QD systems, self-assembled structures show many advantageous features for spin
dynamics. In contrast to, e.g., 
gate-defined lateral or vertical QDs, they are
optically active, allowing one to apply optical control approach originally
developed for bulk semiconductors \cite{kikkawa98} 
and to use light fields to prepare, detect and control spin states on
very short time scales
\cite{dutt05,greilich06c,Atature2006b,Kroner2008a,Xu2008,Dubin2008,Ramsay2008,%
brunner09,Godden2012a}
(see Refs.~\onlineciteb{gywat10,Ramsay2010,Warburton2013,DeGreve_RPP13,Gao2015}
for a review).
Spin relaxation and dephasing in self-assembled QDs is of particular interest,
since these decoherence phenomena set the ultimate limit on the
functionality of any nanoscopic spin-based devices. Experiments show
exciton spin life times much longer than the recombination times
\cite{mackowski04} and
electron spin relaxation times in the range from nanoseconds
\cite{Pal2009} to  microseconds \cite{Heiss2010} or even milliseconds
\cite{kroutvar04,lu10}, depending on the material systems and experimental
details. The measured spin coherence times are much shorter, on the order of
nanoseconds, which is due to hyperfine-induced dephasing and ensemble
inhomogeneity \cite{greilich06c,Akimov2006a,syperek11}. 

Theoretical description of electron spin flip processes in QDs was initially formulated
for a lateral   
gate-defined GaAs structure both for transitions within the ground state
Zeeman doublet\cite{khaetskii01} and for relaxation from higher energy
levels \cite{khaetskii00}. For those structures, the dominant mechanism of spin relaxation
was shown to be the admixture mechanism due to the Dresselhaus
spin-orbit coupling: an electron 
state with a certain nominal spin orientation has a contribution of
states with inverted 
spin, which makes it possible for the phonons to couple it to the
states with a nominally opposite spin orientation. Due to
the time-inversion symmetry, in the resulting effective carrier-phonon
Hamiltonian for the Zeeman doublet, the terms that are even in the magnetic
field $\bm{B}$ have to vanish, which leads to the characteristic $\sim
B^{5}$ dependence of the spin-flip rate.  Two other mechanisms invoked in
Refs.~\onlineciteb{khaetskii00} and \onlineciteb{khaetskii01}, of much lesser importance for
large lateral dots, had a formal structure of a direct
spin-phonon coupling and were interpreted as the spin-orbit splitting of the
electron spectrum due to the strain field produced by the acoustic
phonons and as the
strain-induced modification of the electron Land\'e  
factor. In later literature, an additional ``ripple mechanism'' has been invoked, related
to the phonon-induced 
motion of the QD interface\cite{woods02,Alcalde2005}.
On the other hand, the generic description of spin-phonon coupling, derived within the
formal \kp approach from the phonon-related contributions to the fundamental spin-orbit
Hamiltonian  \cite{pavlov65a,Pavlov1967}, can also be applied to nanostructures in the
effective mass and envelope function approximations \cite{Alcalde2004,Romano2008,Wang2012}.

Based on these results, the most common  theoretical approach followed
in numerous studies, including those devoted to the electron relaxation in self-assembled
structures
\cite{woods02,Westfahl2004,Cheng2004,Alcalde2004,Romano2008,%
Zipper2011,Wang2012,Li2014},
is to derive the electron spin
relaxation rates from a simple model of confinement potential within the 
single-band effective mass approximation and the usual Dresselhaus coupling or other
generic spin-orbit coupling terms. In order to
improve the 
accuracy with which the spin-orbit admixtures to the wave functions
are treated, it was proposed\cite{wei12} to use an atomistic pseudopotential theory
for the calculation of wave functions, while the standard carrier-phonon
coupling was used for the transitions, thus yielding a more exact
theory within the admixture paradigm.
The predictions of the admixture model,
in particular the $B^{5}$ dependence of the spin relaxation rate, have
also been invoked in the interpretation of experimental results that
in fact showed such behavior \cite{lu10,kroutvar04}, apparently
confirming the universal character of those theoretical conclusions.

As the relative strengths of various spin relaxation channels strongly
depend on the system parameters, in particular on the energetic
separation of the excited states, there seems to be no reason for a
particular ordering of these channels to hold universally. Moreover, 
in self-assembled systems, a single-band effective mass approach 
is just the simplest approximation. Applying a more general
multi-band \kp theory in the standard
envelope-function approach\cite{winkler03} not only offers
quantitatively 
more accurate wave functions but also allows one to systematically
include spin-orbit couplings and strain fields. In this
way, all the channels of spin relaxation can be included on the same
footing. The standard quasi-degenerate
perturbation theory (L\"owdin elimination of the valence bands) yields an effective
electron Hamiltonian. This allows one 
to relate the electron spin-flip channels proposed in the literature
to particular terms in the well-established
\kp Hamiltonian as well as to verify the
predictions based on various mechanisms, estimate the effective
constants and assess the relative importance of various couplings for
the electron spin-flip process in a self-assembled QD.

In this paper, we present the results of
\kp modeling of electron spin relaxation in
InGaAs/GaAs self-assembled QDs. First, we classify the terms
responsible for spin-flip processes at the level of an 8-band theory
and show that spin relaxation between the Zeeman sub-levels of the
ground state is dominated by the admixture mechanism induced by shear strain and
valence-band deformation potentials. By perturbatively reducing the model to an effective
description of the 
conduction band, we show that this mechanism corresponds to a strain-dependent
anisotropic contribution to the electron $g$-factor that leads to spin mixing.

The paper is organized as follows. In Sec.~\ref{sec:model} we define the model of the QD
system. In the central Sec.~\ref{sec:results} we present, compare and interpret the
results for spin relaxation via various channels. Sec.~\ref{sec:conclusions} concludes the paper.

\section{Model} 
\label{sec:model}

We consider a flat-bottom lens-shaped, self-assembled InAs QD placed in a GaAs matrix, assuming
a uniform composition of 100\% InAs inside the QD and the wetting layer (WL). The base radius of
the dot is 12 nm and the height is 4.2 nm, while the height of the WL is 0.6 nm.
The system is placed in a magnetic field oriented in the growth direction. 

The electron wave functions are obtained by diagonalizing the 8-band
\kp Hamiltonian in the envelope function approximation \cite{winkler03,Willatzen2009}. 
The model includes the kinetic terms up to
the second order both within the bands and in the band-off-diagonal blocks coupling the
conduction and valence bands.
We account for the strain within the continuous elasticity
approach\cite{pryor98b} in the linear order. 

In the block notation the Hamiltonian has the form
\begin{equation}\label{H-block}
H=\left(\begin{array}{ccc}
H_{\mathrm{6c6c}} & H_{\mathrm{6c8v}} & H_{\mathrm{6c7v}} \\
H_{\mathrm{8v6c}} & H_{\mathrm{8v8v}} & H_{\mathrm{8v7v}} \\
H_{\mathrm{7v6c}} & H_{\mathrm{7v8v}} & H_{\mathrm{7v7v}} \\
\end{array}\right),
\end{equation}
where the blocks refer in the standard way to the lowest conduction band (cb, 6c), the $j=3/2$ valence band (vb, 8v) and the $j=1/2$ (spin-orbit
split-off) vb (7v). Here and in the following we use the
notation of Ref.~\onlineciteb{winkler03}. The corresponding blocks 
are explicitly given by\cite{winkler03,eissfeller11} 
\begin{subequations}
\begin{align}
H_{\mathrm{6c6c}} & 
                          = E_{\mathrm{c}} +V_{\mathrm{p}}            +a_{\mathrm{c}}\tr\epsilon \nonumber \\ 
                  & \quad + \frac{\hbar^{2}}{2m_{0}}  \left( k_{x}  A'_{c} k_{x}
                    +  \frac{i}{2}  k_{[x}g' k_{y]} \sigma_{z} + \cp \right),
  \label{6c6c}
\end{align}
\begin{align}
H_{\mathrm{8v8v}} & 
                          = E_{\mathrm{v}} -\frac{\hbar^{2}}{2m_{0}} \left \{
                          k_{x} \gamma'_{1}  k_{x} - 2 
                            \left (  J^{2}_{x}  - \frac{1}{3} J^{2}  \right )
                          k_{x} \gamma'_{2} k_{x}  \right . \nonumber\\ 
                        &\quad -   \{J_{x},J_{y}\} k_{\{x} \gamma'_{3} k_{y\}}    
                          + \cp   \bigg \} \nonumber\\
                        &\quad + \frac{1}{2\sqrt{3}}  \left[ \{ J_{x},J^{2}_{y} -
                          J^{2}_{z}  \}  \{ C_{k},k_{x} \}   + \cp  \right] \nonumber\\
                          &\quad +a_{\mathrm{v}} \tr\epsilon - b_{\mathrm{v}} \left[ 
                          \left(   J^{2}_{x}  - \frac{1}{3} J^{2} \right)\epsilon_{xx} + \cp \right] \nonumber\\ 
                        & \quad - \frac{d_{\mathrm{v}}}{\sqrt{3}} \left[ \{J_{x},J_{y}\}
                          \epsilon_{xy}  + \cp \right] \nonumber\\
                       &\quad -i \frac{\hbar^{2}}{m_{0}} \left[ k_{[x} \kappa' k_{y]}
                         J_{z} + k_{[x} q k_{y]} J^{3}_{z} + \cp \right ],
  \label{8v8v}
\end{align}
\begin{align}
H_{\mathrm{7v7v}} &= E_{\mathrm{v}} + V_{\mathrm{p}} + a_{\mathrm{v}} \tr\epsilon \nonumber\\
                  &\quad -\Delta_{0} - \frac{\hbar^{2}}{2m_{0}} (k_{x} \gamma'_{1} k_{x} + \cp) \nonumber\\
                  &\quad  -i \frac{\hbar^{2}}{m_{0}} \left[   k_{[x} \kappa' k_{y]}
                    \sigma_{z} + \cp \right ] \nonumber\\ 
                    & \quad - \left ( \mu_{B} B_{z} \sigma_{z} + \cp \right ),
  \label{7v7v}
\end{align}
\begin{align}
H_{\mathrm{6c8v}} & 
                          = \sqrt{3}\bm{T}\cdot\widetilde{\kk} P + i\frac{\sqrt{3}}{2}(T_{x} 
                          k_{\{y} B_{\mathrm{8v}}^{+} k_{z\}} +\cp) \nonumber \\
                        & \quad +\frac{\sqrt{3}}{2}
                         \left [ (T_{xx}-T_{yy}) \left (\frac{2}{3} k_{z}
                          B_{\mathrm{8v}}^{-} k_{z} \right . \right . \nonumber\\ 
                        & \quad  \left . - \frac{1}{3} k_{x} B_{\mathrm{8v}}^{-} k_{x} -
                          \frac{1}{3} k_{y} B_{\mathrm{8v}}^{-} k_{y} \right )\nonumber \\ 
                        & \quad-T_{zz}(k_{x} B_{\mathrm{8v}}^{-} k_{x} - k_{y}
                          B_{\mathrm{8v}}^{-} k_{y}) \bigg ]\nonumber \\ 
                        &\quad +i\sqrt{3}C_2(T_{x}\epsilon_{yz}+\cp), 
                          \label{6c8v}
\end{align}
\begin{align}
H_{\mathrm{6c7v}} & 
                          = -\frac{1}{\sqrt{3}}\bm{\sigma}\cdot\widetilde{\kk} P 
                          -\frac{i}{2\sqrt{3}}(\sigma_{x}  k_{\{y}B_{\mathrm{7v}}
                    k_{z\}} +\cp)\nonumber \\
	                    &\quad -i\frac{1}{\sqrt{3}}C_2(\sigma_{x}\epsilon_{yz} + \cp),   
                    \label{6c7v}
\end{align}
\begin{align}
H_{\mathrm{8v7v}} & = 
                          -\frac{\hbar^{2}}{2m_{0}} \left \{ 
                          -6  (T^{\dagger}_{xx} k_{x} \gamma'_{2} k_{x} + \cp
                          ) \right . \nonumber\\ 
                        & \quad   -6  (T^{\dagger}_{xy} k_{\{x} \gamma'_{3} k_{y\}} + \cp ) \big \} \nonumber\\ 
                        & \quad - i \frac{ \sqrt{3}}{2}  \left ( T^{\dagger}_{yz}
                         \{ C_{k},k_{x} \} + \cp \right ) \nonumber\\
                        & \quad -3 b_{\mathrm{v}} \left( T^{\dagger}_{xx} \epsilon_{xx} +
                          \cp \right) -  \sqrt{3} d_{\mathrm{v}} \left( 2 T^{\dagger}_{xy}
                          \epsilon_{xy} + \cp \right) \nonumber\\ 
                         & \quad   - i \frac{3 \hbar^{2}}{2 m_{0}} 
                           \left[  k_{[x} \kappa' k_{y]} T^{\dagger}_{z} + \cp \right ]\nonumber\\ 
                           & \quad - 3 \left ( \mu_{B} B_{z} T^{\dag}_{z} +
                             \cp \right ). 
 \label{8v7v}
\end{align}
\end{subequations}
Here $\{\mathcal{O}_{1}, \mathcal{O}_{2}\}
=\mathcal{O}_{1}\mathcal{O}_{2}+\mathcal{O}_{2}\mathcal{O}_{1}$,  
$k_{\{i}\mathcal{O} k_{j\}} =
 k_{i} \mathcal{O} k_{j} + k_{j} \mathcal{O} k_{i}$, $k_{[i} \mathcal{O} k_{j]} = k_{i} \mathcal{O} k_{j}
 - k_{j} \mathcal{O} k_{i}$ for any operators $\mathcal{O}$, $\mathcal{O}_{1}$,
 $\mathcal{O}_{2}$; 
c.p. stands for the cyclic permutation of indices;
$E_{\mathrm{c}}$ and $E_{\mathrm{v}}$ are the cb and vb edges, respectively
($E_{0}=E_{\mathrm{c}}-E_{\mathrm{v}}$ is the fundamental band gap in a bulk crystal),
$\Delta_{0}$ is the spin-orbit parameter;
$\kk=-i\nabla+e\bm{A}/\hbar$, where $\bm{A}$ is the vector potential
of the magnetic field $\bm{B}$;
$\widetilde{\kk}=\kk (\mathbb{I}-2\epsilon)$; 
$\epsilon$ is the strain tensor;
$V_{\mathrm{p}}$ is the piezoelectric potential including piezoelectric polarization up to
second-order terms in structural strain \cite{bester06,Migliorato2014} with the parameters
taken from Ref.~\onlineciteb{caro15};   
$m_{0}$ is the free electron mass; 
$\gamma_{i}'$ are the Luttinger parameters with removed contributions
from the $\Gamma_{6}$ cb, 
\begin{equation*}  
  \gamma'_{1} =  \gamma_{1} -  \frac{E_{P}}{3 E_{0} + \Delta_{0}},   \;
  \gamma'_{2,3} =  \gamma_{2,3} - \frac{1}{2} \frac{E_{P}}{3 E_{0} + \Delta_{0}};
\end{equation*} 
$\mu_{B}$  is the Bohr magneton;
$q$ is the anisotropic contribution to the bulk $g$-factor in the Luttinger Hamiltonian;
$\sigma_{i}$ are the Pauli matrices; $J_{i}$ are matrices of the $j=3/2$
representation of angular momentum; $T_{i}$ are matrix representations
of a vector operator between $j=1/2$ and $j=3/2$ 
 states, i.e., $T_{x/y}=-(T^{(1)}_{+1}\mp T^{(1)}_{-1})/\sqrt{2}$, $T_{z}=T^{(1)}_{0}$,
with the matrix elements of the spherical components $T^{(1)}_{q}$
given in terms of the Clebsch-Gordan coefficients 
$\langle j_{1}j_{2};m_{1}m_{2}|jm\rangle$ by the 
Wigner-Eckart theorem, 
$\langle m |T^{(1)}_{q} | m'\rangle = -\sqrt{2/3}\langle 3/2,m';1,q|1/2,m\rangle$,
for $m=\pm 1/2$, $m'=-3/2,\ldots,3/2$; and 
$T_{ij} = T_{i}J_{j}+T_{j}J_{i}$;
$A'_{c}$, $g'$ and $\kappa'$  are given by\cite{winkler03}
\begin{align*} 
 A'_{c} &\equiv \frac{m_{0}}{m'} =  \frac{m_{0}}{m^{*}} - \frac{2}{3} \frac{E_{P}}{E_{0}} -
                \frac{1}{3} \frac{E_{P}}{E_{0}+\Delta_{0}}, \\
  g'&= 2, \quad
 \kappa'= -\frac{1}{3} \left ( \gamma'_{1} - 2 \gamma'_{2} - 3 \gamma'_{3} + 2 \right ).
\label{g-prim}
\end{align*}
In order to avoid $A'_{c} < 0$, which would result in a non-elliptical system \cite{birner11},
we use $E_{P} = (m_{0}/m^{*} - 1 ) E_{0}(E_{0}+\Delta_{0})/(E_{0}+2\Delta_{0}/3)$, which guarantees $A'_{c} = 1$.
In view of the inconsistency of the reported values of $q$\cite{winkler03,lawaetz71}, we
follow Ref.~\onlineciteb{eissfeller12} and use the perturbative formula  
$q = (2/9) E_{Q}\Delta'_{0}/[E'_{0}(E'_{0}+\Delta'_{0})]$,
where $E_{Q}$, $E'_{0}$ and $\Delta'_{0}$ are the $14$-band $\kp$
parameters\cite{winkler03};
then $P=\hbar(E_{P}/2m_{0})^{1/2}$.
In numerical calculations we use the gauge-invariant discretization scheme\cite{andlauer08}
for the covariant derivative.
The material parameters used in our $\kp$ calculations are given in
Table~\ref{tab:param}.

\begin{table}
	\begin{tabular}{lccl}
		\toprule
		& GaAs\tabtopspacer & InAs & Interpolation for $\mathrm{In_{x}Ga_{1{\minus}x}As}$\\
		\hline
		$E_{\mathrm{v}}$ & $0.0$~eV\tabtopspacer & $0.21$~eV & linear\\
		$E_{\mathrm{0}}$ & $1.519$~eV & $0.417$~eV & $0.417x{\plus}1.519(1{\minus}x){\minus}0.477x(1{\minus}x)$\\
		$E'_{\mathrm{0}}$ & $4.488$~eV & $4.390$~eV & linear\\
		$E_{\mathrm{Q}}$ & $17.535$~eV & $18.255$~eV & linear\\
		\multirow{2}{*}{$m^{*}$} & \multirow{2}{*}{$0.0665m_{0}$} & \multirow{2}{*}{$0.0229m_{0}$} & $  [ 0.0229x {\plus} 0.0665(1{\minus}x)$\\    
		& & & ${\minus}0.0091x(1{\minus}x)] m_{0}$ \\
		$\Delta$ & $0.341$~eV & $0.39$~eV& $0.39x{\plus}0.341(1{\minus}x){\minus}0.15x(1{\minus}x)$ \\ 
		$\Delta'_0$ & $0.171$~eV & $0.25$~eV& linear \\ 
		$a_{\mathrm{c}}$ & $-7.17$~eV & $-5.08$~eV & ${\minus}5.08x{\minus}7.17(1{\minus}x){\minus}2.61x(1{\minus}x)$\\
		$a_{\mathrm{v}}$ & $1.16$~eV & $1.00$~eV & linear\\
		$b_{\mathrm{v}}$ & $-2.0$~eV & $-1.8$~eV & linear\\
		$d_{\mathrm{v}}$ & $-4.8$~eV & $-3.6$~eV & linear\\ 
		$\gamma_{\mathrm{1}}$ & $6.98$ & $20.0$ & $1/ \left [ (1{\minus}x)/6.98 {\plus} x/20.0 \right ]$ \\
		$\gamma_{\mathrm{2}}$ & $2.06$ & $8.5$ & $1/\left [ (1{\minus}x)/8.5 {\plus} x/2.06 \right ]$\\
		$\gamma_{\mathrm{3}}$ & $2.93$ & $9.2$ & $1/\left [ (1{\minus}x)/9.2 {\plus} x/2.93 \right]$\\  
          $C_{2}$ & \multicolumn{2}{c}{$3.3$~eV} &  \\
		\toprule
	\end{tabular} 
	\caption{\label{tab:param}Material parameters used in the
          calculations. After Refs.~\onlineciteb{Vurgaftman2001,winkler03}, except for
          $C_{2}$, which is extracted from the measurement data in Ref.~\onlineciteb{dyakonov86a}.} 
\end{table}

Coupling to acoustic
phonons is included in the standard way in the long-wavelength limit, by taking into account the 
phonon-related contribution to the strain tensor in the \kp Hamiltonian, expressing it in
terms of the phonon-induced displacements and quantizing the latter. In 
addition, piezoelectric coupling is taken into account by performing the same procedure on
the strain terms entering the Hamiltonian via induced piezoelectric fields. The
off-diagonal piezoelectric couplings are discussed in the
\hyperref[sec:appendix]{Appendix} and are shown to give negligibly small contribution,
hence we do not include them in the Hamiltonian. 
The Zeeman splitting at 10~T is 1.065~meV, which corresponds to the wave numbers of 0.31
and 0.58~nm$^{-1}$ for longitudinal and transverse phonons, respectively. This
corresponds to 2.8\% and 5.2\% of the  Brillouin zone, respectively, thus justifying the
long wave length approximations, as well as the linear dispersion model.

\section{Results and interpretation}
\label{sec:results}

In this section, we present the results for the transition rate between the states forming the Zeeman doublet of the electron
ground state, obtained from the 8-band \kp calculations. First, in Sec.~\ref{sec:types}, we discuss the general division of the
spin-flip channels into two classes. Next, in Sec.~\ref{sec:contributions} we present the numerical
results for the spin-flip rates resulting from various channels. The dominant channel
is then related to an effective term in a reduced conduction band Hamiltonian in
Sec.~\ref{sec:relation}.

\subsection{Admixture and spin-phonon mechanisms}
\label{sec:types}

The purpose of our analysis is to assess the quantitative importance of
various spin-flip mechanisms and to identify the leading ones. First, however, let us note
that the direct carrier-phonon coupling is spin-conserving and the original conduction-band block
of the multi-band \kp Hamiltonian is spin-diagonal, which precludes any spin-flip
transitions unless the coupling to valence bands is taken into account via a perturbation
theory (L\"owdin partitioning\cite{lowdin51}). The resulting spin-flip mechanisms that may appear as
higher-order perturbations in the effective description of conduction-band electrons can
be of two kinds\cite{khaetskii00,khaetskii01}. The first type are \textit{admixture mechanisms}, where
the spin transition is due to admixture of states with opposite spin, which makes it
possible for phonons to couple two such states\cite{pikus84,khaetskii00,khaetskii01}. The second
class are \textit{spin-phonon mechanisms}, resulting from symmetry-lowering phonon-related strain fields, which,
combined with the spin-orbit coupling in the valence bands, lead to direct ``spin-phonon''
terms in the effective conduction-band Hamiltonian\cite{Roth1960,pikus84}. 

These two kinds
of processes appearing in the effective conduction-band description can be mapped back to
the 8-band \kp model and used to classify the results of the multi-band modeling. For this
purpose, let us split the effective conduction-band Hamiltonian into the spin-diagonal
zeroth-order part $H_{0}$, the spin-conserving electron-phonon coupling $V_{0}$, and the
perturbative correction resulting from decoupling of the valence band. The
latter contains strain-dependent terms, kept up to the linear order, and is 
not diagonal in spin states. The instantaneous strain field, represented by the strain
tensor $\epsilon$, is composed of the static strain due to the system inhomogeneity
$\epsilon^{(\mathrm{s})}$ and the phonon-induced contribution
$\epsilon^{(\mathrm{ph})}$, which leads to decomposition of the perturbative correction
into two spin-non-diagonal terms with generic forms, respectively, 
\begin{displaymath}
H_{1}=\sum_{i=0}^{3}\sum_{jk=1}^{3}\alpha_{ijk}\sigma_{i}\epsilon_{jk}^{(\mathrm{s})},\quad
V_{1}=\sum_{i=0}^{3}\sum_{jk=1}^{3}\alpha_{ijk}\sigma_{i}\epsilon_{jk}^{(\mathrm{ph})},
\end{displaymath}
where $\sigma_{0}$ is the unit matrix and $\sigma_{i}$ for $i=1,2,3$ are the Pauli
matrices. By diagonalizing $H=H_{0}+H_{1}$ and computing phonon-induced transition rates
resulting from $V=V_{0}+V_{1}$ one obtains  in principle all the spin-conserving and
spin-flipping transitions in the system. To the leading order, however, the latter can be
induced either by a combination of $H_{1}$ and $V_{0}$ (admixture mechanisms) or $H_{0}$
and $V_{1}$ (spin-phonon mechanisms). It is therefore clear that the distinction between
these two classes of processes can be traced back to the place where phonons are coupled
into the 8-band model: in the conduction-band block of the multi-band \kp Hamiltonian
(admixture mechanism) or in the other blocks, mapped onto the conduction band upon
L\"owdin perturbative decoupling (spin-phonon mechanisms).

\subsection{Contributions to the spin-flip rate}
\label{sec:contributions}

\begin{figure}[tb]
\begin{center}
\includegraphics[width=\columnwidth]{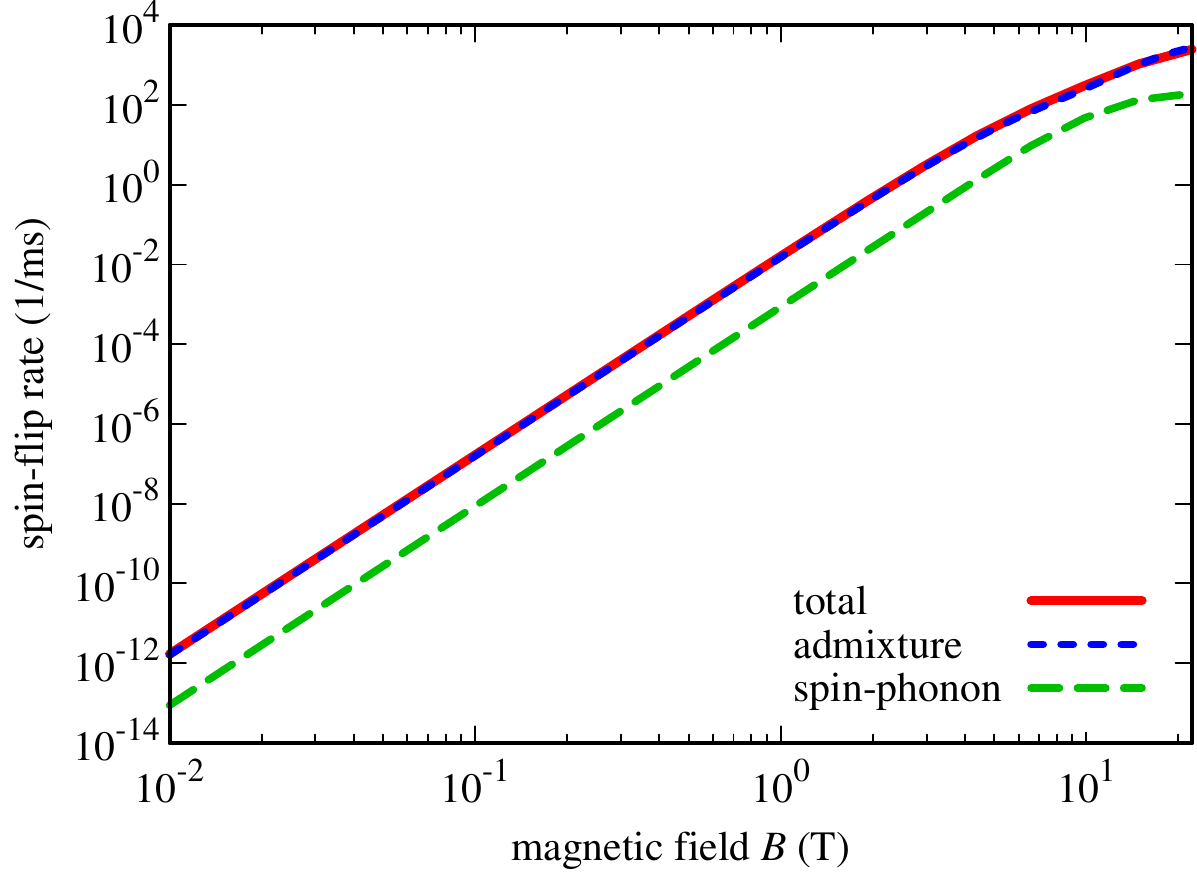} 
\end{center}
\caption{\label{fig:general}(Color online) The total relaxation rate (solid red line) compared to the rates
  due to the admixture (dotted blue ) and spin-phonon (dashed green) mechanisms only, as a
  function of the magnetic field.}   
\end{figure}

In Fig.~\ref{fig:general}, we show the total spin-flip rate (solid red line), as well as
the rates resulting from admixture and spin-phonon mechanisms only (dotted blue and dashed
green lines, respectively) as a function of the magnetic field $B$. The admixture mechanisms
dominate over the other by over an order of magnitude in the whole range of magnetic fields. Both
contributions scale as $B^{5}$ up to about 10~T and at stronger fields the $B$-dependence
saturates. The two contributions are almost exactly additive (see Tab.~\ref{tab:values}
for explicit values).

\begin{table}
  \centering
  \begin{tabular}{ll|ll}
  	\toprule
    \multicolumn{4}{c}{spin-flip mechanisms and rates (s$^{-1}$)}\tabtopspacer \\
    \hline
    \multicolumn{4}{c}{\textbf{total rate \quad\quad 16.31}}\tabtopspacer \\
    \hline
    \textbf{total admixture}\tabtopspacer & \textbf{15.42} & \textbf{total spin-phonon} & \textbf{0.874}  \\ \hline
    $d_{\mathrm{v}}$ strain\tabtopspacer & 1.385 & $d_{\mathrm{v}}$ phonons & 1.169 \\
    off-diag strain & 3.477 & off-diag phonons & 0.0237 \\ 
    \textbf{$d_{\mathrm{v}}$ + off-diag strain} & \textbf{15.36} \\
    Dresselhaus & 0.238 \\
    $C_{2}$ off-diag strain & 0.175 & $C_{2}$ off-diag phonons & 0.0353 \\
    ``none'' & 0.185 \\
    \toprule
  \end{tabular}
  \caption{\label{tab:values}Numerical values of the spin-flip transition rate at $B=1$~T for
    individual mechanisms and selected combinations of mechanisms.}
\end{table}
 
\begin{figure}[tb]
\begin{center}
\includegraphics[width=\columnwidth]{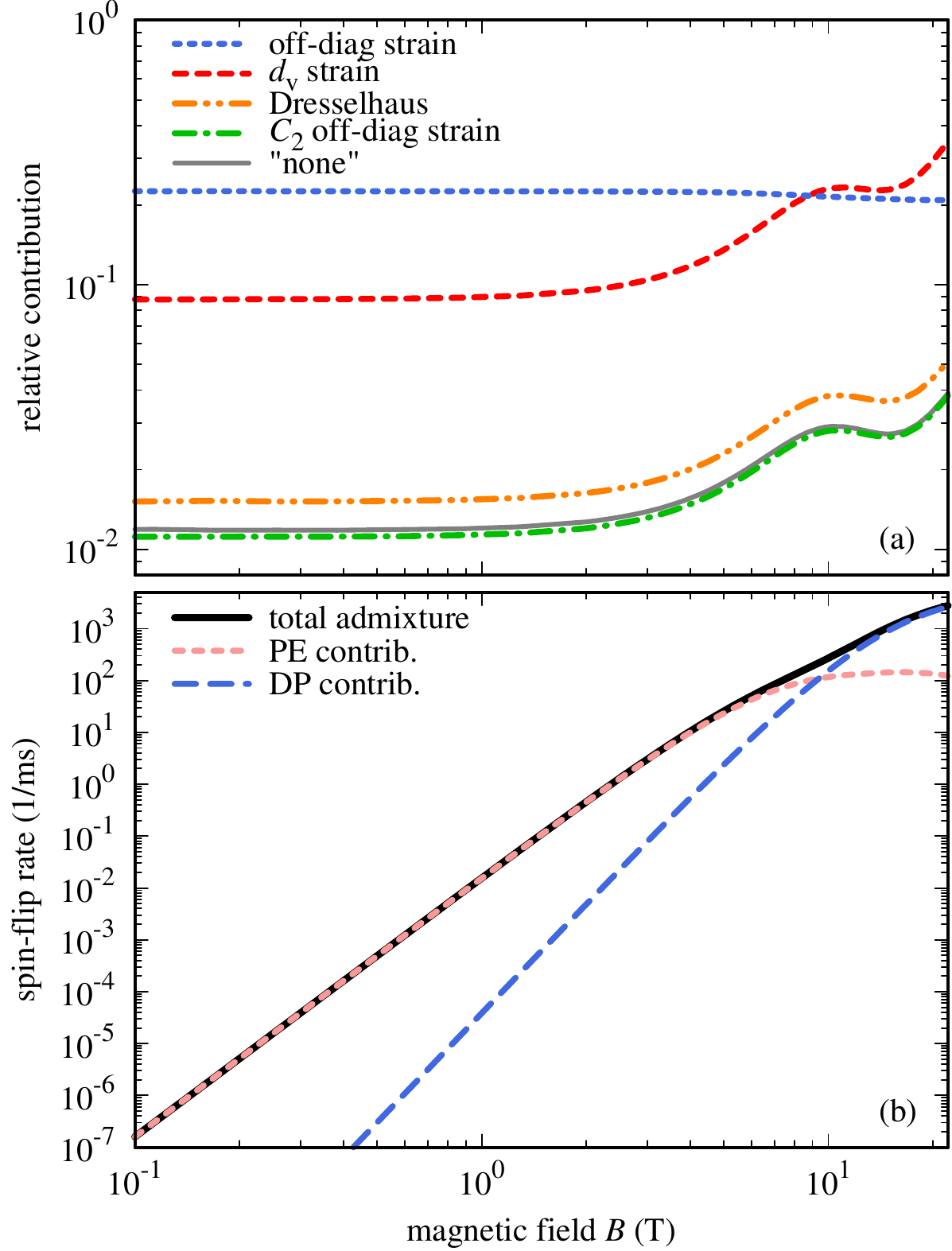} 
\end{center}
\caption{\label{fig:admix}(Color online) (a) Relative contributions to the admixture mechanism from various couplings
  in the valence band and band-off-diagonal blocks of the \kp Hamiltonian as a function of
  the magnetic field. Each line shows the ratio of the spin flip rate with only one mechanism
  turned on to the total admixture-induced rate shown in Fig.~\ref{fig:general}.  The thin
  gray solid line  shows the results for all the explicit terms turned off.
 (b) Contributions of the couplings to different phonon branches as a function of the
 magnetic field. }  
\end{figure}

The spin-non-conserving admixture can originate either from the Dresselhaus spin-orbit
coupling, represented by quadratic terms in $H_{\mathrm{6c8v}}$ and $H_{\mathrm{6c7v}}$
(which is the dominant mechanism in large QDs \cite{khaetskii01}), or from various terms
in the valence band,  
reflecting spin-orbit couplings in a nanostructure, where the crystal symmetry is broken
on the mesoscopic level by composition inhomogeneity and strain. In order to determine
the dominant contribution in a self-assembled QD, we have studied the spin-flip
transition rate for individual contributions to the admixture channel. To this end, we
have identified terms in the \kp Hamiltonian that lead to spin relaxation via the admixture
mechanism (with carrier-phonon coupling via diagonal terms in the conduction
band only) and calculated the rate with all
these terms switched off in our computational model, except for a single one.
The results are shown in
Fig.~\ref{fig:admix}(a), where we plot the contributions relative to the total
admixture-induced rate. 
One can see that no single contribution dominates the overall
rate. The two most-important ones stem from the shear strain terms 
that are linear in both momentum and strain in the band-off-diagonal blocks of the \kp
Hamiltonian $H_{\mathrm{6c8v}}$ and $H_{\mathrm{6c7v}}$ (dashed blue line, labeled
``off-diag strain''  in Fig.~\ref{fig:admix}(a)) 
and from the terms in the valence-band blocks $H_{\mathrm{8v8v}}$ and
$H_{8v7v}$ proportional to the deformation potential $d_{\mathrm{v}}$ (dashed red line). 
In order to facilitate quantitative comparison, the rates at $B=1$~T
are listed in Tab.~\ref{tab:values}. The rate obtained when both the dominant
channels are turned on is nearly equal to the total rate for admixture mechanisms, while
the other mechanisms yield less than 1\% of the rate. Note that these two major rates are
not additive; in fact, their joint effect is larger than expected even assuming
constructive interference of transition amplitudes (which is indeed the case, see
Sec.~\ref{sec:relation} for more insight). The reason is the large impact of the strain
terms in  $H_{\mathrm{6c8v}}$ and $H_{\mathrm{6c7v}}$ on the electron $g$-factor: with these terms on, the
Zeeman splitting increases by 77\% (from 61 to 108~\si{\micro}eV), which enhances relaxation due
to growing phonon spectral density at higher frequencies.

The contribution of the remaining channels is very small and leads altogether to a 0.4\%
correction to the result. This is mostly due to a small Rashba contribution from the overall valence-band-edge
inhomogeneity, piezoelectric field in the valence band, and 
interfaces, which cannot be switched off in our numerical model and remains after
all the other explicit couplings are removed; this is represented by the thin solid grey line
labeled ``none'' in Fig.~\ref{fig:admix}(a). Actually, the effect of interfaces is
dominant: switching the piezoelectric field in the valence band off reduces this
contribution by 3\% only. The familiar Dresselhaus coupling
($B_{\mathrm{7v}}$ and $B_{\mathrm{8v}}$ terms in  $H_{\mathrm{6c8v}}$ and
$H_{\mathrm{6c7v}}$) adds some 50\% to this Rashba spin-flip rate. The strain terms
proportional to the $C_{2}$  deformation potential contribute negligibly and turn out to interfere
destructively with the Rashba part, slightly decreasing the total rate when switched on. 
These results are in strong contrast to what was found for large QDs in a simple
single-band confinement 
model (corresponding to unstrained, gated QDs) \cite{khaetskii01},  where the single
Dresselhaus coupling was shown to dominate. 

In Fig.~\ref{fig:admix}(b), we have compared the contributions from the
deformation-potential(DP, short-dashed orange line)
and piezoelectric (PE, dashed blue line) couplings to phonons in $H_{\mathrm{6c6c}}$, which may lead to spin
flip in the admixture mechanisms. Up to about 5~T the total rate (black solid lines), as well as its
individual components (not shown in the plot) are nearly entirely due to the
piezoelectric coupling. As a result, all the rates scale with the magnetic
field as $B^{5}$. Deformation-potential coupling produces a $B^{7}$ contribution that is
negligible at low and moderate fields but becomes important from about 10~T.

\begin{figure}[tb]
\begin{center}
\includegraphics[width=\columnwidth]{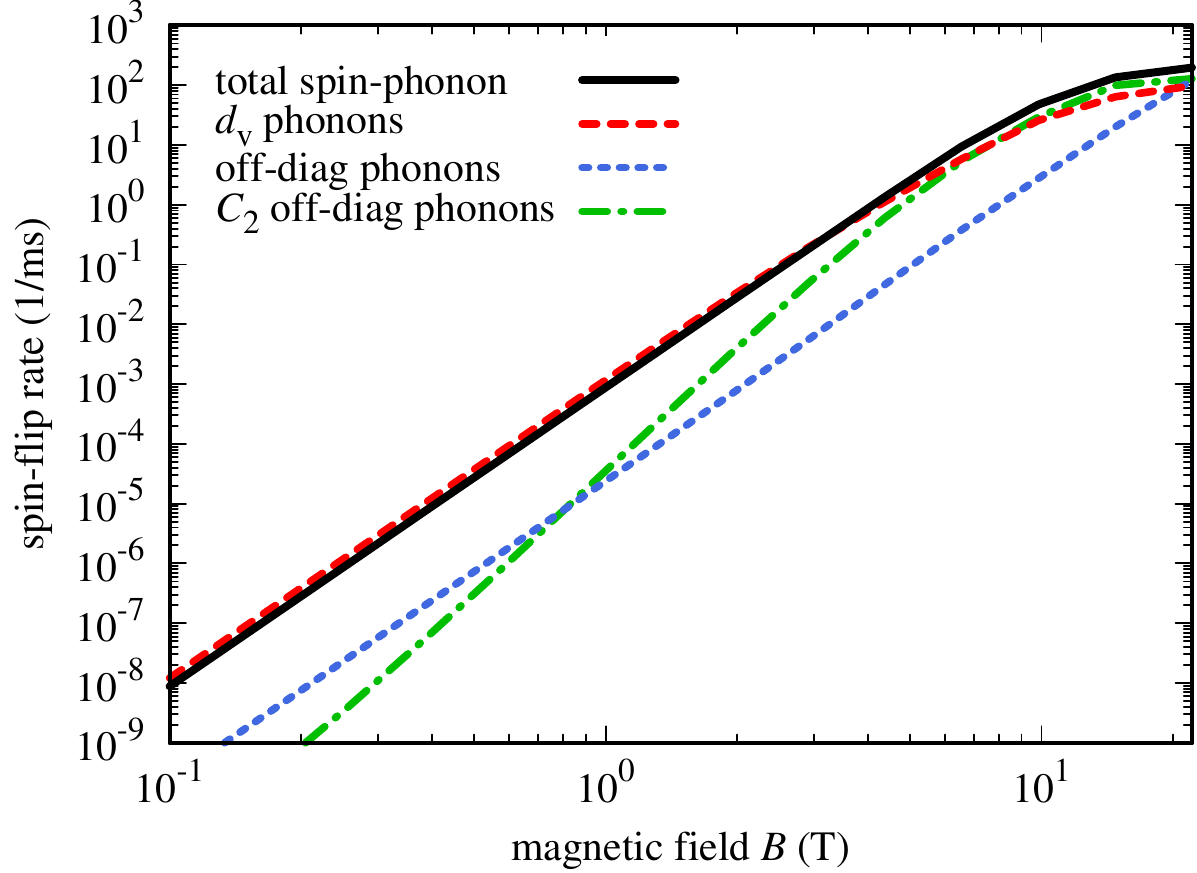}
\end{center}
\caption{\label{fig:spinphon}(Color online) Contributions to the spin-phonon mechanism
  from various phonon couplings 
  in the valence band as a function of the magnetic field.}  
\end{figure}

In Fig.~\ref{fig:spinphon} we show selected contributions to the spin-phonon
mechanism. Here, the total rate due to this mechanism is clearly dominated by one coupling:
the terms proportional to $d_{\mathrm{v}}$ in the valence blocks (with some destructive
interference with the other  channels). The couplings in off-diagonal blocks
have much less importance here at low and moderate fields. However, while the dominant
coupling shows a $B^{5}$ behavior up to about 5~T, the
$C_{2}$ coupling grows as $B^{7}$ in the range of fields shown (it has a $B^{5}$ to
$B^{7}$ crossover at about 0.2~T) and becomes relatively important at field magnitudes of
a few Tesla.

\subsection{Interpretation in terms of an effective Hamiltonian 
for the conduction band}  
\label{sec:relation}

We shall now relate the dominant spin-flip contributions to the effective strain-related
corrections to the electron Hamiltonian, most of which have been known in different
contexts in literature.

Perturbative decoupling of the valence band leads to a correction to the conduction band
Hamiltonian, the relevant part of which can be written as\cite{Mielnik-Pyszczorski2017} 
\begin{equation}
  \label{H2-T}
H_{\mathrm{eff}}  = S \mathcal{D}^{-1}S^{\dag}, 
\end{equation}
where $S =
\sum_{jl}k_{j}(\delta_{jl}-\epsilon_{jl})\tcal_{l}P+iC_{2}\sum_{j}\epsilon_{j}\tcal_{j}$. Here,
$\mathcal{T}_{i}=\sqrt{3}T_{i}\oplus(-1/\sqrt{3})\sigma_{i}$ (a $2\times 6$ matrix),
$\epsilon_{x}\equiv\epsilon_{yz}$ etc. (by cyclic permutations),
$\mathcal{D}=\chi_{\mathrm{c}}\mathbb{I}_{6\times 6} -\widetilde{H}_{\mathrm{v}}$ represents
the structure of the valence band, with
$\widetilde{H}_{\mathrm{v}}$ approximating the $6\times6$ valence-band block of the
\kp Hamiltonian, renormalized by the L\"owdin procedure (the details can be found in 
Ref.~\onlineciteb{Mielnik-Pyszczorski2017}, but are not relevant here), and
$\chi_{\mathrm{c}}$ is the local conduction band edge (neglecting magnetic contributions,
the conduction band block $H_{\mathrm{6c6c}}$ [Eq.~\eqref{6c6c}] is proportional to unit
matrix and can be represented by a scalar function $\chi_{\mathrm{c}}$).  

In order to extract the admixture contribution of the terms proportional to $d_{\mathrm{v}}$,
we write $\mathcal{D}=\mathcal{D}_{0}-H_{\mathrm{v}}^{(\mathrm{dv})}$,
where $\mathcal{D}_{0}=\mathrm{diag}(\Eg,\Eg,\Eg,\Eg,\Eso,\Eso)$ is a diagonal
approximation to $\mathcal{D}$ accounting for the 
major position-dependent band-edge shifts due to composition and strain (heavy-light hole
splitting could be included here as well, in order to slightly improve accuracy, but is
neglected for simplicity), and 
\begin{displaymath}
H_{\mathrm{v}}^{(\mathrm{dv})} = -\sqrt{3}d_{\mathrm{v}}\es_{xy} \left(
  \begin{array}{cc}
    \frac{1}{3}\left\{ J_{x},J_{y}\right\}  & 2T_{xy}^{\dag} \\
    2T_{xy} & 0
  \end{array} 
\right) +\mathrm{c.p.} 
\end{displaymath}
is the part of the valence-band block proportional to $d_{\mathrm{v}}$. Here, the blocks of
the matrix notation refer to the $\Gamma_{\mathrm{8v}}$ (heavy- and light-hole) and
$\Gamma_{\mathrm{7v}}$ (spin-orbit split-off) bands. 
Substituting this approximate form of $\mathcal{D}$ to Eq.~\eqref{H2-T}, one gets for the
relevant part of the conduction band Hamiltonian
\begin{eqnarray}\label{dv}
\lefteqn{H_{\mathrm{eff}}^{(\mathrm{dv})}=}\\ 
&  &  \kk\cdot\bm{\sigma} \frac{2\sqrt{3} P^{2}d_{\mathrm{v}}}{\Eg(\Eso)}
\left(T_{xy}\es_{xy}+\mathrm{c.p.}\right) 
\bm{T}^{\dag}\!\cdot\kk +\hc \nonumber\\
& & -\kk\cdot\bm{T} \frac{\sqrt{3}P^{2}d_{\mathrm{v}}}{\Eg^{2}}
\left(\left\{J_{x},J_{y}\right\}\es_{xy}+\mathrm{c.p.}\right) 
\bm{T}^{\dag}\!\cdot\kk,
\nonumber
\end{eqnarray}
where we omitted the strain-related contributions to $S$, in order to keep the
result linear in strain.
The spin-dependent contributions result from the antisymmetric parts of the two terms in
Eq.~\eqref{dv}, defined as 
$[T_{j}\mathcal{O}T_{l}]_{\mathrm{as}}\equiv(1/2)(T_{j}\mathcal{O}T_{l}-T_{l}\mathcal{O}T_{j})$
and 
$[\sigma_{j}\mathcal{O}T_{l}]_{\mathrm{as}}
\equiv(1/2)(\sigma_{j}\mathcal{O}T_{l}-\sigma_{l}\mathcal{O}T_{j})$
for any operator $\mathcal{O}$. Substituting the explicit forms of the matrices one finds
\begin{align*}
\left[T_{x} \left\{J_{x},J_{y}\right\}  T_{z}\right]_{\mathrm{as}} & = 
4\left[\sigma_{x} T_{xy}  T_{z}\right]_{\mathrm{as}}
=\frac{i}{3}\sigma_{x}, \\
\left[T_{z} \left\{J_{x},J_{y}\right\} T_{y}\right]_{\mathrm{as}} & = 
4\left[\sigma_{z} T_{xy} T_{y}\right]_{\mathrm{as}}
=\frac{i}{3}\sigma_{y},
\end{align*}
with other non-zero terms obtained by asymmetry and by cyclic permutation of indices. 
Neglecting the non-commutativity of $k_{j}$ with $P$, $d_{\mathrm{v}}$, and $\Eg$ and
using the relations $[k_{j},k_{l}]=-i(e/\hbar)\sum_{n}\varepsilon_{jln}B_{n}$,
$eP^{2}/\hbar=\mu_{\mathrm{B}}E_{\mathrm{P}}$, one obtains 
\begin{equation}\label{eff-dv}
H_{\mathrm{eff}}^{(\mathrm{dv})} = 
\frac{1}{2}\mu_{\mathrm{B}}\bm{B}\delta\hat{g}^{(\mathrm{dv})}\bm{\sigma},
\end{equation}
where $\delta\hat{g}$ is a tensor with elements
\begin{displaymath}
\delta g_{jl}^{(\mathrm{dv})} = 
\frac{2}{\sqrt{3}}\frac{E_{P}d_{\mathrm{v}}\Delta_{\mathrm{SO}}}{\Eg^{2}(\Eso)}\es_{jl}
\end{displaymath}
and $E_{P}=2mP^{2}/\hbar^{2}$.
This strain-induced mixing term is known in literature \cite{VanVleck1940,Roth1960} and
can be interpreted as a correction $\delta g_{jl}^{(\mathrm{dv})}$ to the electron Land\'e
tensor \cite{khaetskii00,khaetskii01}. 

The second largest contribution, which is due to strain terms in the off-diagonal block of
the \kp Hamiltonian, enters the effective conduction-band Hamiltonian via strain terms
proportional to $P$ in $S$. We now approximate
$\mathcal{D}\approx\mathcal{D}_{0}$, which yields, up to linear order in strain, 
\begin{equation}\label{Pe}
H_{\mathrm{eff}}^{(\mathrm{od})} = -\sum_{jnl}
k_{j}P\es_{jn}\tcal_{n}\mathcal{D}_{0}^{-1}
\tcal_{l}Pk_{l} + \hc
\end{equation}
Again, only the antisymmetric part yields a spin-dependent term. Using the explicit form of
$\tcal_{j}$, one finds 
\begin{equation}\label{as-part}
\left[ \tcal_{n}\mathcal{D}_{0}^{-1}\tcal_{l}\right]_{\mathrm{as}} =
-\frac{i}{3}\frac{\Delta_{\mathrm{SO}}}{\Eg(\Eso)}\sum_{m}\varepsilon_{nlm}\sigma_{m},
\end{equation}
from which one gets 
\begin{displaymath}
H_{\mathrm{eff}}^{(\mathrm{od})} = 
\frac{1}{2}\mu_{\mathrm{B}}\bm{B}\delta\hat{g}^{(\mathrm{od})}\bm{\sigma},
\end{displaymath}
where
\begin{displaymath}
\delta g_{jl}^{(\mathrm{od})} = 
-\frac{2}{3}\frac{E_{P}\Delta_{\mathrm{SO}}}{\Eg(\Eso)}\es_{jl}.
\end{displaymath}
This term has exactly the same structure as the previous one, which explains why
the two contributions to spin admixture interfere constructively (note that
$d_{\mathrm{v}}$ is negative).

The most important mechanism in the class of direct spin-phonon couplings at low and
moderate fields, stemming from the $d_{\mathrm{v}}$ 
terms, can be mapped onto an effective conduction band Hamiltonian exactly in the same way
as the $d_{\mathrm{v}}$ term discussed above. The static strain is replaced by $\eph$, expressed in
terms of lattice displacements, and expanded in plane-wave modes. Upon
quantization of the latter, one obtains an effective spin-phonon coupling Hamiltonian,
discussed already in Refs.~\onlineciteb{khaetskii00} and \onlineciteb{khaetskii01}, which describes spin
transitions due to phonon-induced dynamical anisotropic modulation of the $g$-factor.

The spin-phonon contribution proportional to $C_{2}$ appears via terms linear in both $C_{2}$
and strain in Eq.~\eqref{H2-T},
\begin{equation*}
H_{\mathrm{eff}}^{(C2)} = iC_{2}\sum_{jl}
\epsilon_{j}\tcal_{j} \mathcal{D}_{0}^{-1} \tcal_{l}Pk_{l} + \hc
\end{equation*}
Using Eq.~\eqref{as-part}, one immediately finds
\begin{equation}\label{eff-C2}
H_{\mathrm{eff}}^{(C2)} = \frac{1}{3}\sum_{jlm}\varepsilon_{jlm}\left\{
\frac{C_{2}\Delta_{\mathrm{SO}}P}{\Eg(\Eso)}\epsilon_{j},k_{l} \right\}\sigma_{m}.
\end{equation}
A similar term was derived in the context of the calculation of energy spectrum of strained
semiconductors \cite{bir61,pikus84,dyakonov86a}. It was discussed as a spin-phonon
coupling in the analysis of spin relaxation channels in large QDs
\cite{khaetskii00,khaetskii01}, where it was shown to be much less effective than the
``$d_{\mathrm{v}}$'' spin-phonon channel discussed above. Interestingly, in a simple model
of the ``particle-in-a-box'' confinement with real ground state wave functions, the effective
Hamiltonian $H_{\mathrm{eff}}^{(C2)}$ leads to $B^{5}$ dependence of the spin-flip
rate\cite{khaetskii01},  
while the numerical values from the 8-band \kp theory yield a crossover to $B^{7}$
dependence already below 1~T, as discussed in Sec.~\ref{sec:contributions}.

Another spin-phonon term in the effective Hamiltonian appears from off-diagonal
piezoelectric couplings to phonons (see \hyperref[sec:appendix]{Appendix}). This term is non-zero only in an
inhomogeneous system but even here its effect is small.

\section{Conclusions}
\label{sec:conclusions}

In this paper we have presented the results of theoretical modeling of spin-flip
relaxation between Zeeman sub-levels of a single electron in a 
self-assembled InAs/GaAs QD. By analyzing the spin relaxation process with 8-band envelope-function
\kp theory, we have identified individual spin-flip channels
divided into two classes: admixture and spin-phonon mechanisms. We have shown that the
former dominates, like in large unstrained dots, although not so overwhelmingly (95\% of
the total rate). However, in sharp contrast to the latter,
the dominant channel of spin relaxation in strained self-assembled QDs is spin admixture
induced by symmetry-breaking shear strain, which accounts for 99.6\% of the total
admixture-induced rate. The dominant processes can be mapped onto two different effective
spin-phonon terms in a conduction band (effective mass) Hamiltonian that interfere and
interplay in a non-trivial way in producing the total spin-flip rate. 

The most important practical consequence of our findings is that the dominant
contribution to spin relaxation in self-assembled QDs relies on the particular
distribution of shear strain in the structure and, therefore, cannot be modeled by a
unique standard term in the Hamiltonian. This is in sharp contrast to larger, unstrained
dots, where spin relaxation is dominated by the Dresselhaus coupling easily accounted for by
the well-known generic spin-orbit term in the Hamiltonian. 

The second observation that we find important is that in magnetic fields up to about 5~T
the rates of all the important spin-flip channels, both admixture and spin-phonon, are
proportional to $B^{5}$. Therefore, this simple characteristics cannot be used as a key to
distinguishing the dominant spin-flip mechanisms in an experiment.

Finally, identifying the dominant spin-flip mechanism as being due to strain suggests
that considerable enhancement of spin life time may be possible in structures with reduced
strain. This is consistent with the results concerning an impurity-bound electron
\cite{Linpeng2016}, which 
is a strain-free system, where the dominating spin-flip channels are related to direct
spin-phonon and Dresselhaus SO couplings. 
  
\acknowledgments
The authors acknowledge support from the
Polish National Science Centre under Grant No.~2014/13/B/ST3/04603 (AM-P, KG, PM) and Grant 
No.~2014/14/M/ST3/00821 (MG).
Calculations have been carried out using resources provided by
Wroclaw Centre for Networking and Supercomputing (\url{http://wcss.pl}), Grant No. 203.

\appendix*

\section{Off-diagonal piezoelectric couplings}\label{sec:appendix}

In this Appendix, we derive the general structure of the off-diagonal piezoelectric
carrier-phonon couplings and estimate the resulting spin-phonon terms in the effective
Hamiltonian for conduction band electrons.

The strain due to phonons, written in the coordinate representation with respect to the
electron and in the second quantization with respect to phonons, has the form
\begin{displaymath}
	\epsilon_{ij} \lr{ \rr } = \sumql \epsilon_{ij} ^{ (\bm{q},\lambda) } e^{i\qq\cdot\rr},
\end{displaymath}
where 
\begin{multline}\label{eq:strain-ph}
\epsilon_{ij}^{ (\bm{q},\lambda) }=
	-\hf \sqrt{\frac{\hbar}{2 \rho V \wql}}  \\
	\times \Lr{\, \lr{ \eql }{\!}_i \,q_j + \lr{ \eql }{\!}_j \, q_i } \lr{ \bql + \bmqld }.
\end{multline}
Here $V$ is the normalization volume of the phonon system, $\eql=-\emqlc$ is the mode
polarization, and $\bqld,\bql$ are phonon creation and annihilation operators. 
The resulting piezoelectric potential in a zincblende crystal is then
\begin{displaymath}
V(\rr)=i\sumq v^{(\qq)}  e^{i\qq\cdot\rr},
\end{displaymath}
where
\begin{displaymath}
 v^{(\qq)} = 2 \mathcal{E}_{\mathrm{p}}\frac{1}{q^{2}}\sum_{\lambda}
\left( q_{x}\epsilon_{yz}^{ (\bm{q},\lambda) }+\cp \right).
\end{displaymath}
The above equation is correct for an inhomogeneous system in the long wave length limit,
when the small-scale details become irrelevant and the system can be approximated by a
virtual uniform medium characterized by a 
constant $\mathcal{E}_{\mathrm{p}}=ee_{14}/\epsilon_{0}\epsilon_{s}$, which should be
close to the GaAs matrix value of 1.4~eV/nm.

In the envelope-function approach, one separates the mesoscopic length scales
(coarse-grained position $\RR$) from the atomic scales (position $\bm{\xi} $ within a unit
cell). It is assumed that material parameters vary only on the mesoscopic scales. The
matrix elements of a multi-band \kp Hamiltonian at a position $\RR$ are then obtained as
matrix elements of the original Hamiltonian between the Bloch functions $u_{\mu},u_{\nu}$
corresponding to the two bands $\mu,\nu$, calculated over one unit cell (u.c.) of volume
$v$, located at $\RR$.
Writing
$\rr=\RR+\bm{\xi}$, one obtains the contribution of the piezoelectric coupling to the
matrix elements of the \kp Hamiltonian
\begin{displaymath}
V_{\mu\nu}=i\sumq v^{(\qq)}  e^{i\qq\cdot\RR}
\left\langle \mu \left| e^{-\qq\cdot \bm{\xi} } \right| \nu \right\rangle_{\mathrm{u.c.}},
\end{displaymath}
where 
\begin{displaymath}
\left\langle \mu\left| \mathcal{O}(\bm{\xi})\right| \nu
\right\rangle_{\mathrm{u.c.}}
= \frac{1}{v}\int_{\mathrm{u.c.}}\mathrm{d}^{3} \bm{\xi} \, u_{\mu}^{*}(\bm{\xi}) 
\mathcal{O}(\bm{\xi}) u_{\nu}(\bm{\xi}).
\end{displaymath}
In a mesoscopic structure, the magnitude of $\qq$ for phonons that are efficiently coupled to
confined carriers is effectively limited to the range $q \lesssim 1/l \ll 1/a$, where $l$
is the spatial extension of the envelope function and $a$ is the lattice constant. Therefore,
$\qq\cdot\bm{\xi} \ll 1$ and the exponent can be expanded in series, 
\begin{displaymath}
e^{i\qq\cdot\bm{\xi}} 	\simeq 1 + i\qq\cdot\bm{\xi} - \frac{\lr{\qq\cdot\bm{\xi}}^2}{2}.
\end{displaymath}
The zeroth-order term is diagonal due to the orthogonality of Bloch functions and for
each of the bands reproduces the standard piezoelectric carrier-phonon coupling. The
higher-order terms contribute to inter-band couplings, for which the zeroth-order term vanishes.

With the known composition of Bloch functions in terms of atomic orbitals one can relate
the required matrix elements to those between angular momentum eigenstates. Here, we will
take the standard assumption that the cb states are $s$-type and the vb states are purely
$p$-type. Then, due to parity, the linear term in the expansion contributes only to the off-diagonal cb-vb
block of the \kp Hamiltonian. Denoting 
$\langle 1/2,1/2,\mathrm{cb} |\xi_{x} | 3/2,3/2,\mathrm{vb}\rangle = 
d_{1}/\sqrt{2}$ one finds from the Wigner-Eckart theorem
\begin{equation*}
\left\{ \left\langle\mu \left|\bm{\xi} \right| \nu \right\rangle_{\mathrm{u.c.}} \right\}_{\mu,\nu} 
=\sqrt{3} d_1 \bm{\tcal},	
\end{equation*}
where $\mu$ and $\nu$ run through the two conduction and six  valence bands, respectively.
 Hence, the resulting first-order Hamiltonian term is
\begin{align}\label{H-PE1}
H^{(\mathrm{PE,1})}
	&= \sqrt{3} d_{1} \bm{\mathcal{E}}\cdot \bm{\tcal} + \hc,
\end{align}
where 
\begin{displaymath}
\bm{\mathcal{E}}(\RR) = -\nabla V(\RR) = \sumq  v^{(\qq)} \bm{q} e^{i\qq\cdot\RR}
\end{displaymath}
is the piezoelectric field. 

The quadratic term has non-vanishing matrix elements only between valence band
states. Denoting $\langle 3/2,3/2,\mathrm{vb} |\xi_{x}\xi_{y} | 3/2,-1/2,\mathrm{vb}\rangle =
-i d_{2}/\sqrt{3}$, with $d_{2}$ real, one finds the relevant part of the valence-band
piezoelectric perturbation
\begin{equation}\label{H-PE2}
H^{(\mathrm{PE,2})}=\frac{i}{6}d_{2}\frac{\partial \mathcal{E}_{y}(\RR)}{\partial x}\left(
  \begin{array}{cc}
    \left\{ J_{x},J_{y}\right\}  & 6T_{xy}^{\dag} \\
    6T_{xy} & 0
  \end{array} 
\right) +\mathrm{c.p.},
\end{equation}
where we neglected terms proportional to $q_{i}^{2}$ that do not induce spin relaxation.

In order to assess the effect of the linear term \eqref{H-PE1} on the electron spin-flip
processes, we go back to Eq.~\eqref{H2-T}, where we extend $S\to S'=S+S_{\mathrm{pe}}$
with $S_{\mathrm{pe}}=H^{(\mathrm{PE,1})}$. From the resulting terms we again select the
spin-dependent antisymmetric part, according to Eq.~\eqref{as-part}. The resulting effective
Hamiltonian for the conduction band can be written in two equivalent forms
\begin{align}\label{eff-PE1}
H_{\mathrm{eff}}^{\mathrm{(PE,1)}} & = -\frac{i}{\sqrt{3}}
\frac{\Delta_{\mathrm{SO}}Pd_{1}}{\Eg(\Eso)}\bm{\sigma}\cdot \left(
\bm{\mathcal{E}\times \kk}\right) +\hc \nonumber\\
&=\frac{1}{\sqrt{3}}\left[\bm{\mathcal{E}}\times\nabla
\frac{\Delta_{\mathrm{SO}}Pd_{1}}{\Eg(\Eso)}\right]\cdot\bm{\sigma},
\end{align}
where we used the fact that $\bm{\mathcal{E}}$ is longitudinal.
The first equation represents the effective Hamiltonian in the usual Dresselhaus form with
the piezoelectric field as the symmetry-breaking factor. The final equation shows
explicitly that the block-off-diagonal terms contribute to electron spin-phonon coupling only in an
inhomogeneous system. 
By comparing  Eq.~\eqref{eff-PE1} with Eq.~\eqref{eff-C2} one can see that the overall
magnitude of the piezoelectric spin-flip term is reduced by a factor
$d_{1}\mathcal{E}_{\mathrm{p}}/C_{2}$. In GaAs $d_{1} = 0.11$~nm (estimated in a
model of hydrogen-like orbitals with equal 
distribution of wave functions between the anion and the
cation\cite{Chekhovich2011,Clementi1963}). Hence,
$d_{1}\mathcal{E}_{\mathrm{p}}/C_{2}=0.047$ and  we expect the resulting rate
(proportional to the square of the coupling) to be at least three
orders of magnitude lower than that resulting from the $C_{2}$ coupling, which is small itself.

The effective Hamiltonian corresponding to Eq.~\eqref{H-PE2} is constructed by closely
following the derivation of Eq.~\eqref{eff-dv}. One obtains the analogous term
\begin{equation*}
H_{\mathrm{eff}}^{(\mathrm{PE,2})} = 
\frac{1}{2}\mu_{\mathrm{B}}\bm{B}\delta\hat{g}^{(\mathrm{PE,2})}\bm{\sigma},
\end{equation*}
with
\begin{displaymath}
\delta g_{jl}^{(\mathrm{PE,2})} = 
-\frac{2}{3}\frac{E_{P}d_{2}\Delta_{\mathrm{SO}}}{\Eg^{2}(\Eso)}
\frac{\partial \mathcal{E}_{j}}{\partial x_{l}}, \quad j\neq l.
\end{displaymath}
Comparison to Eq.~\eqref{eff-dv} shows that the piezoelectric term is smaller by a factor
$\mathcal{E}_{\mathrm{p}}d_{2}q_{0}/d_{\mathrm{v}}=1.1\times 10^{-4}$, where we used the estimate $d_{2}=
9.3\times 10^{-3}$~nm$^{2}$ (obtained in the same way as $d_{1}$ above) and
$q_{0}=g\mu_{\mathrm{B}}B/(\hbar c) \approx 0.04$~nm$^{-1}$ 
is the resonant wave vector for a transition between Zeeman sub-levels. It follows that
this term is negligible.


\end{document}